\documentclass{sig-alternate}
\usepackage{array}
\usepackage{graphicx}
\usepackage{url}
\usepackage{xcolor}
\usepackage[numbers]{natbib} 

\newcommand{\bhv}{\hat{\mathbf{v}}}
\newcommand{\rmi}{\mathrm{i}}
\newcommand{\rmd}{\mathrm{d}}
\newcommand{\sfH}{\mathsf{H}}

\begin{document}

\permission{}
\copyrightetc{Center for Advanced Research Computing Technical Report CARC-2015-174}

\title{Performance Analysis of an Astrophysical Simulation Code on the Intel Xeon Phi Architecture}

\numberofauthors{3} 
\author{
\alignauthor
Vahid Noormofidi\\
       \affaddr{Department of Electrical and Computer Engineering}\\
       \affaddr{University of New Mexico}\\
       \affaddr{Albuquerque, New Mexico}\\
       \email{vahid@unm.edu}
\alignauthor
Susan R. Atlas\\
       \affaddr{Department of Physics and Astronomy}\\
       \affaddr{University of New Mexico}\\
       \affaddr{Albuquerque, New Mexico}\\
       \email{susier@unm.edu}
\alignauthor
Huaiyu Duan\\
       \affaddr{Department of Physics and Astronomy}\\
       \affaddr{University of New Mexico}\\
       \affaddr{Albuquerque, New Mexico}\\
       \email{duan@unm.edu}
}

\maketitle
\begin{abstract}
We have developed the astrophysical simulation code XFLAT to study
neutrino oscillations in supernovae. XFLAT is designed to utilize
multiple levels of parallelism through MPI, OpenMP, and SIMD
instructions (vectorization). It can run on both CPU and Xeon Phi
co-processors based on the Intel Many Integrated Core Architecture
(MIC). We analyze the performance of XFLAT on configurations with CPU
only, Xeon Phi only and both CPU and Xeon Phi. We also investigate the
impact of I/O and the multi-node performance of XFLAT on the Xeon
Phi-equipped Stampede supercomputer at the Texas Advanced Computing
Center (TACC). 
\end{abstract}


\section{Introduction}
At the end of its life, the core of a massive star collapses under its
own gravity and releases a gigantic amount of energy which blows up the
rest of the star in a (core-collapse) supernova
\cite{Woosley:2005yv}. Supernovae are 
essential to the chemical evolution of the universe. 
During a supernova chemical elements such as carbon, oxygen,
gold and uranium are ejected into the inter-stellar medium in which
new generations of stars such as our sun are born.

It turns out
that 99\% of the energy of a supernova is carried away
by a group of nimble particles called neutrinos and their
anti-particles or antineutrinos. There are three kinds
or ``flavors'' of neutrinos: electron-flavor, mu-flavor and tau-flavor
neutrinos. It has been well established by various experiments that
a neutrino of one flavor can mutate into another flavor during
propagation \cite{Agashe:2014kda}. This phenomenon is known as neutrino
(flavor) oscillation 
or transformation. Because neutrinos play an influential role in supernovae,
neutrino oscillations can also be important in, e.g., supernova
dynamics and the synthesis of chemical elements.

In a typical supernova approximately $10^{58}$ neutrinos are emitted
in just tens of seconds. These neutrinos form a dense neutrino medium
surrounding the collapsed stellar core which becomes a neutron
star. Under appropriate conditions the neutrino medium can experience
collective oscillations in which neutrinos of different energies and
initial flavors, emitted from different positions and propagating in
different directions, can all become coupled \cite{Duan:2010bg}.
Computing collective neutrino oscillations is a very challenging
problem, and it has been done only in some greatly simplified models.

\begin{figure}
\centering
\includegraphics[width=\columnwidth]{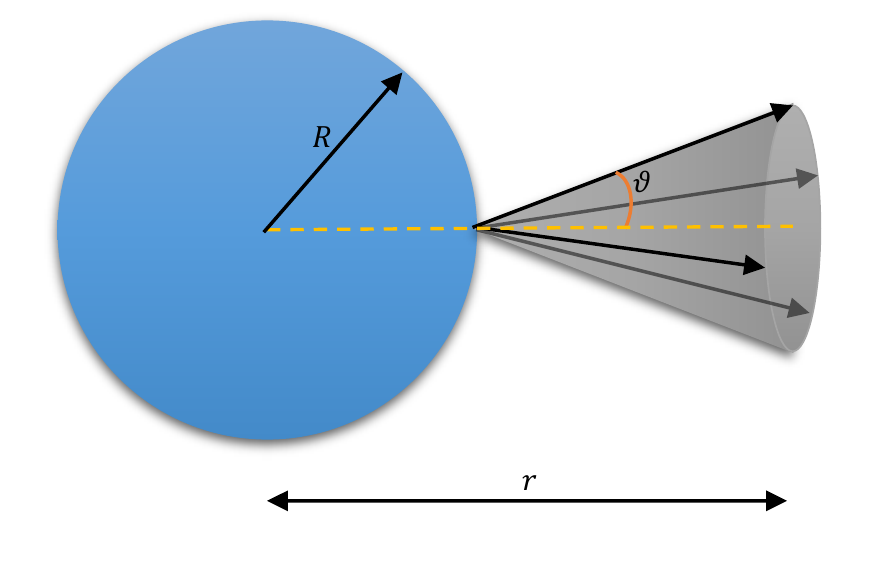}
\caption{Geometry of the (neutrino) bulb model.}
\label{fig:bulb}
\end{figure}

The most sophisticated supernova model by now in which collective neutrino
oscillations can be computed self-consistently is the (neutrino) \emph{bulb
model} (Fig.~\ref{fig:bulb}) \cite{Duan_Simulation_2006}. For
simplicity, the bulb model assumes 
spherical symmetry about the center of the neutron star and 
no time dependence. The neutron star in this model is
described as a sphere of radius $R$ from the surface of which
neutrinos are emitted. 
In the bulb model the flavor
quantum state $\psi_\alpha(\bhv,E;r)$ of a neutrino at radius $r$ is
determined by the Schr\"odinger equation
\begin{align}
\rmi\frac{\rmd}{\rmd r}\psi_\alpha(\bhv,E;r)
= (\sfH_0 + \sfH_\nu)\cdot \psi_\alpha(\bhv,E;r),
\label{eq:eom}
\end{align}
where $\alpha$, $\bhv$ and $E$ are the initial flavor, (the unit vector
of) the propagation direction, and the energy of the neutrino,
respectively, $\sfH_0$ is the Hamiltonian in the absence of the
neutrino medium, and $\sfH_\nu$ is the neutrino potential because of
the ambient neutrinos. 
The propagation direction $\bhv$ is fully described by the polar angle
$\vartheta$ between $\bhv$ and the radial direction when axial
symmetry about the radial direction is imposed. In the \emph{extended bulb
model} \cite{Mirizzi:2013rla} where axial symmetry is not imposed,
$\bhv$ is determined by both $\vartheta$ and the azimuthal angle
$\varphi$ about the radial axis.
For $n$ neutrino flavors, $\psi$ is a vector of
$n$ complex variables, and $\sfH_0$ and $\sfH_\nu$ are both $n\times n$
Hermitian matrices. 

The difficulty in solving Eq.~\eqref{eq:eom} stems from the neutrino
potential:
\begin{align}
\sfH_\nu = \sum_{\alpha'} \int\!\rmd E'\int\!\rmd\bhv'\,
(1-\bhv\cdot\bhv')[F'\psi'(\psi')^\dagger
-\bar{F}'(\bar\psi')^*(\bar\psi')^T],
\label{eq:Hv}
\end{align}
where the quantities with primes are associated with the ambient
neutrinos, the quantities with bars are associated with
antineutrinos, $\psi'=\psi_{\alpha'}(\bhv',E';r)$, and
$F'=F_{\alpha'}(\bhv',E';r)$ is the number flux of the neutrino.
Eqs.~\eqref{eq:eom} and \eqref{eq:Hv} require the self-consistent solution for the flavor quantum states of all neutrinos simultaneously.

\section{Numerical implementation}
Neutrino oscillations in the neutrino bulb model can be solved
numerically by using, e.g.\ FLAT, which is a numerical code developed
by one of us \cite{Duan_Simulating_2008}. In FLAT the quantum flavor
states $\psi_\alpha(\vartheta,E;r)$ of neutrinos at a given radius $r$
are described by a multi-dimensional object array 
\texttt{psi\_alpha[theta, E]}. At each radius $r$ summations over the elements of this array with different weighting functions are performed to obtain the
neutrino potential $\sfH_\nu$,
and a modified midpoint method is subsequently employed to evolve
$\psi_\alpha(\vartheta,E;r)$ over one radial step. In a typical run
$100-1,000$ discrete energy bins are needed to achieve the desired
energy resolution, and $1,000-10,000$ polar angle ($\vartheta$) bins
are required to 
achieve numerical convergence. In other words, millions to tens of
millions nonlinear differential equations need to be solved
simultaneously in computing neutrino oscillations in the bulb model.
The problem sizes of more realistic models can be significantly
larger. For example, the inclusion of the azimuthal ($\varphi$) dimension in the
extended bulb model can increase the problem size significantly.

Because the flavor quantum states of all neutrinos are required to
compute $\sfH_\nu$, it is desirable that these calculations are run on
as few compute nodes as possible so as to minimize the inter-node communication. At the same time, the larger problem sizes of better supernova models
require greater computing power. This balance can be achieved by
employing accelerators or co-processors such as Graphics Processing Units (GPU) and the Xeon Phi, which in principle have much higher throughput than CPUs. 
To maintain the same code set for both CPU and co-processor we chose to
develop for the Intel Xeon Phi co-processor which is based on the Intel
Many Integrated Core Architecture (MIC).

XFLAT is written in C++ and uses an algorithm similar to that of FLAT.
However, unlike FLAT, which uses only MPI, XFLAT implements three levels of
parallelism. In Fig.~\ref{fig:xflat} we illustrate the high level
structure of XFLAT for the extended neutrino bulb model. 
At the top level, the $\vartheta$ angle bins are distributed among MPI 
nodes which can be either a CPU or a Xeon Phi co-processor. At the
middle level (i.e.\ on a CPU or Xeon Phi) $\vartheta$ angle bins are
dispatched to OpenMP threads. At the 
bottom level (i.e.\ in a thread) the loop over energy bins is
performed using vectorization or Single Instruction with Multiple
Data (SIMD). In order to use SIMD efficiently we use multiple
double-precision floating-point arrays
\texttt{ar\_alpha[E]}, \texttt{ai\_alpha[E]}, etc.\ to represent the real
and imaginary parts of the variables in complex vectors
$\psi_\alpha(\vartheta,\varphi,E;r)$ of the same $\varphi$
and $\vartheta$ and at a given $r$. These arrays are then grouped into
an object element of neutrino beam array \texttt{NBeam[theta,phi]}.

\begin{figure}
\begin{verbatim}
/// Distribute theta bins among MPI nodes.
...
/// Main computation loop.
while (!termination_conditions)
{
  /// Dispatch local theta bins among OpenMP threads.
  #pragma omp parallel for
  for (theta: LOCAL_THETA_COUNT)
  {
    /// Loop over phi bins.
    for (phi: PHI_COUNT)
    {
      /// Loop over energy bins using SIMD.
      #pragma simd
      for (E: ENERGY_COUNT)
      {
        /// Evolve psi over one radial step.
        ...
        /// Perform sums over energy bins.
        ...
      }
      /// Perform sums over phi bins.
      ...
    }
  /// Perform partial sums over local theta bins.
  ...
  }
  /// Exchange partial sums among MPI nodes.
  ...
  /// Compute neutrino potential.
  ...
}
\end{verbatim}
\caption{High-level structure of XFLAT.}
\label{fig:xflat}
\end{figure}

\section{Performance Benchmarks}
We validated XFLAT against FLAT using the bulb model. In this paper we present
XFLAT benchmarks using the \emph{extended} bulb
model with the azimuthal angle ($\varphi$) dependence which is not
supported by FLAT. Most of the benchmarks are presented for a
``standard problem'' with 2
neutrino flavors, 100 energy bins, 10 azimuthal angle ($\varphi$)
bins and 10,000 polar angle ($\vartheta$) bins. This problem size fits on the
memory of a single Xeon Phi.
When both the CPU and Xeon Phi were used in a benchmark, the number of
polar angle bins on the Xeon Phi 
is three times as large as that on the CPU unless stated otherwise.

XFLAT benchmarks were performed on the Stampede supercomputer
at the Texas Advanced Computing Center (TACC)
(see Table~\ref{tab:stampede} for Stampede specifications). 
Double precision was used exclusively in all floating point
calculations. XFLAT and all kernels were
compiled with the Intel C++ compiler v13.0.2
with flags \texttt{-O3 -openmp -xHOST} for CPU and
\texttt{-O3 -openmp -mmic} for Xeon Phi, respectively.
One XFLAT process was run on each CPU and/or Xeon Phi (i.e.\ the
Xeon Phi was run in \emph{native} mode \cite{JeffersReinders201303}).
Each process on the CPU was run with 8 OpenMP threads since
hyperthreading is disabled on Stampede \cite{Stampede}. Each process
on Xeon Phi was run with
244 OpenMP threads to fully utilize its hyperthreading capability.
For MPI-enabled benchmarks the Intel
MPI library v4.1.3.049 was used, and for I/O-enabled benchmarks the
NetCDF v4.3.2 and HDF5 v1.8.13 libraries were used. 

\begin{table}[t]
\centering
\caption{Stampede Dell PowerEdge C8220z compute node specifications \cite{Stampede}.}
\renewcommand{\arraystretch}{1.2}
\begin{tabular}[t]{ p{2cm}  p{5.7cm} } \hline\hline
Component & Specifications \\ \hline
CPU & 2x Xeon E5-2680 (8 cores) @2.7GHz (turbo, @3.5GHz)\\
Memory  & 32GB (8x4GB) quad channel DDR3 @1600MHz\\ 
Disk & 250GB 7.2K RPM SATA\\ 
Co-processor & 2x Xeon Phi SE10P (61 cores) @1.1GHz
8GB GDDR5 @5500MHz \\ \hline\hline
\end{tabular}
\label{tab:stampede}
\end{table}

\begin{figure}[t]
\begin{verbatim}

/// 8/244 OpenMP threads for CPU/Xeon Phi.
#pragma omp parallel for
for (t: NUM_THREADS)
{
  /// Repeat 10 million times.
  for (itr: LOOP_COUNT)
  {
    /// VECTOR_WIDTH is a multiple of 4/8 
    /// for CPU/Xeon Phi.
    #pragma simd
    for (s: VECTOR_WIDTH)
    {
      /// floating point operations
      ...
    }
  }
}
\end{verbatim}%
\caption{High-level structure of the benchmark code for floating
  point performance.}%
\label{fig:benchmark}%
\end{figure}

\subsection{Single-node benchmarks}

\begin{figure}[t]
\centering
\includegraphics[width=\columnwidth]{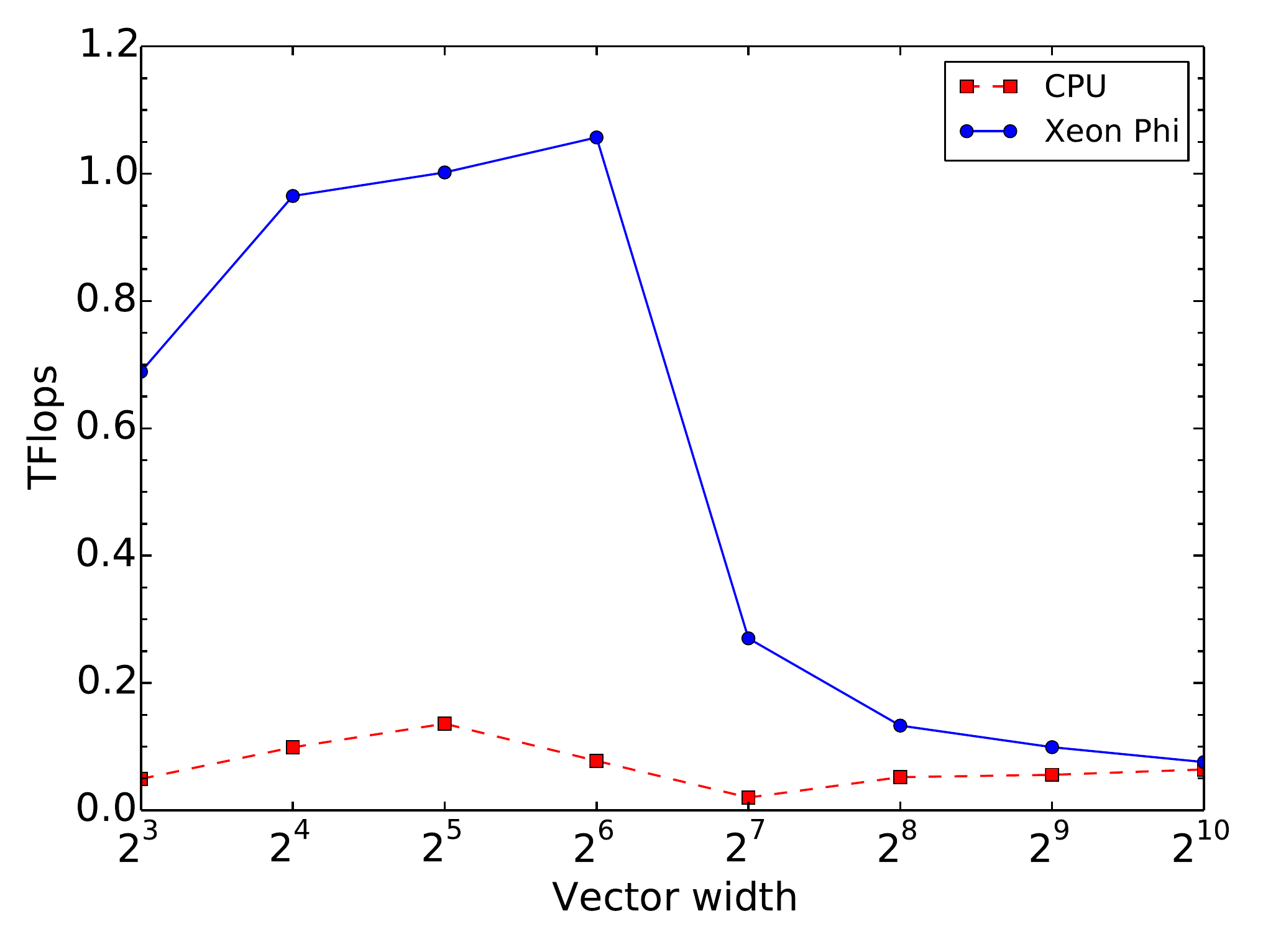}
\caption{Floating point performance of CPU and Xeon Phi.}
\label{fig:flops}
\end{figure}

We first benchmarked the raw performance of the CPU and
Xeon Phi on  
Stampede using a code adapted from \cite{JeffersReinders201303} with the
structure illustrated in 
Fig.~\ref{fig:benchmark}. In the innermost loop of this
code simple floating point operations (i.e.\ additions and
multiplications) are performed on an array or vector of
double-precision (DP) floating-point 
numbers. The widths of the vectors are taken to be a multiple of that of
the SIMD registers of the computing component (256 bits or 4 DP for
the Xeon CPU, and 512 bits or 8 DP for the Xeon Phi). The same vector operations
are repeated 10 million times in the middle loop to maintain data
locality. In the outermost loop all of the hardware threads are utilized
to achieve the best performance. 
The results of these benchmarks with different vector widths in the
innermost loop are shown in Fig.~\ref{fig:flops}. These results
show that
the floating point performance of the Xeon Phi is highly sensitive to the width
of the vector, while the performance of the CPU is relatively stable. The
floating point performance of the Xeon Phi is best when the width
of the DP vectors is 64; in that test, the Xeon Phi ran 10 times as fast as the
CPU. However, the performance of the Xeon Phi degrades substantially as
the vector width increases.

\begin{figure}
\centering
\includegraphics[width=\columnwidth]{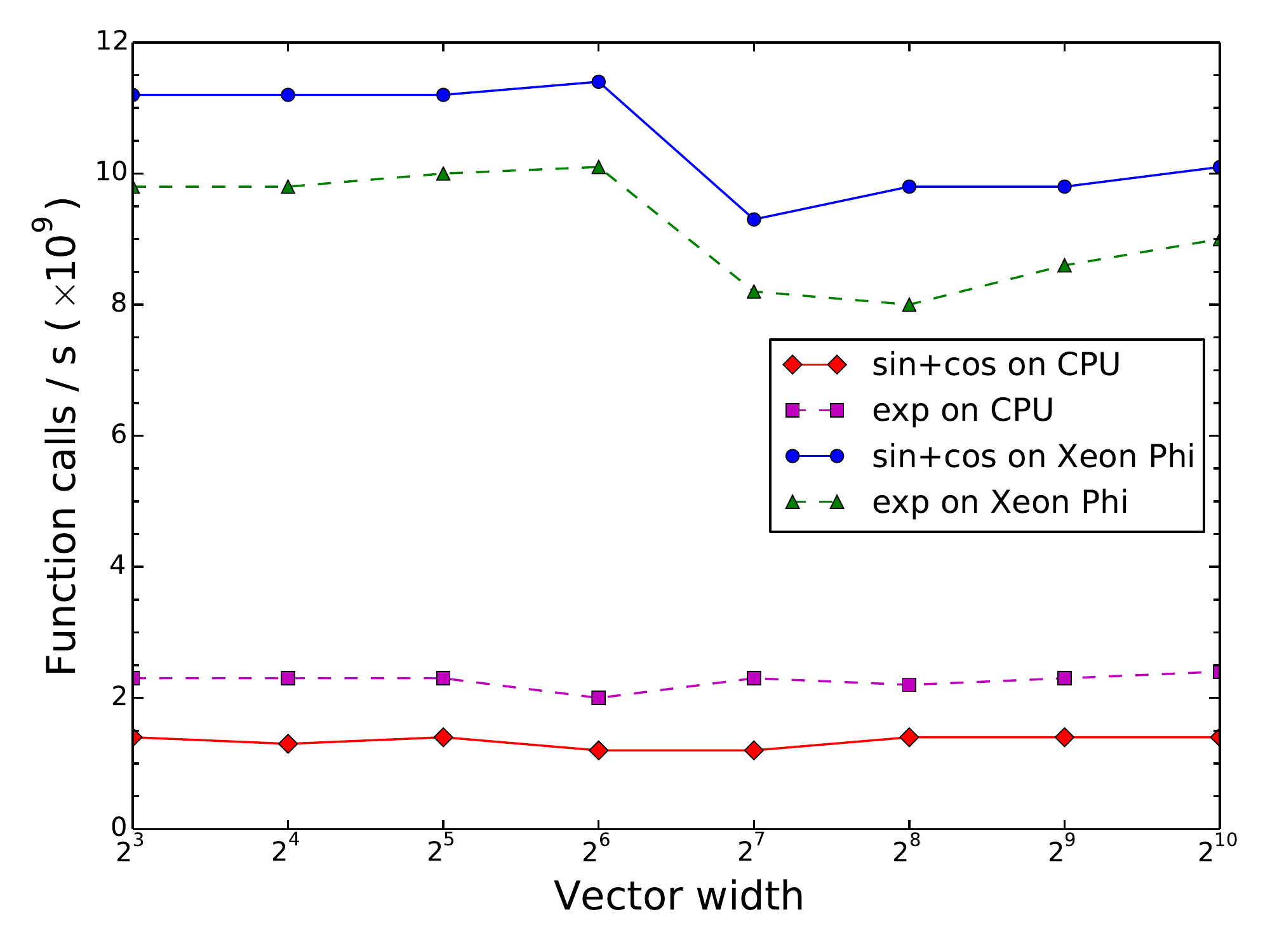}
\caption{Transcendental function performance of CPU and Xeon Phi.}
\label{fig:sin_exp}
\end{figure}

Because XFLAT uses transcendental functions such as \texttt{sin()} and
\texttt{exp()}, we also benchmarked the transcendental function
performance of the CPU and Xeon Phi using a code similar to that of 
Fig.~\ref{fig:benchmark} with the simple floating point operations
replaced by a pair of \texttt{sin()} and \texttt{cos()} functions in
one series of tests, 
and \texttt{exp()} in the other. The results of
these benchmarks are shown in Fig.~\ref{fig:sin_exp}. These results
suggest that the transcendental function performance on the Xeon Phi is
relatively stable against the width of the vectors. For the tests on
\texttt{sin()} and \texttt{cos()} the Xeon Phi ran 6--8 times as fast
as the CPU, but for \texttt{exp()} the Xeon Phi is
only 3--4 times better.

\begin{figure}[t]
\centering
\includegraphics[width=\columnwidth]{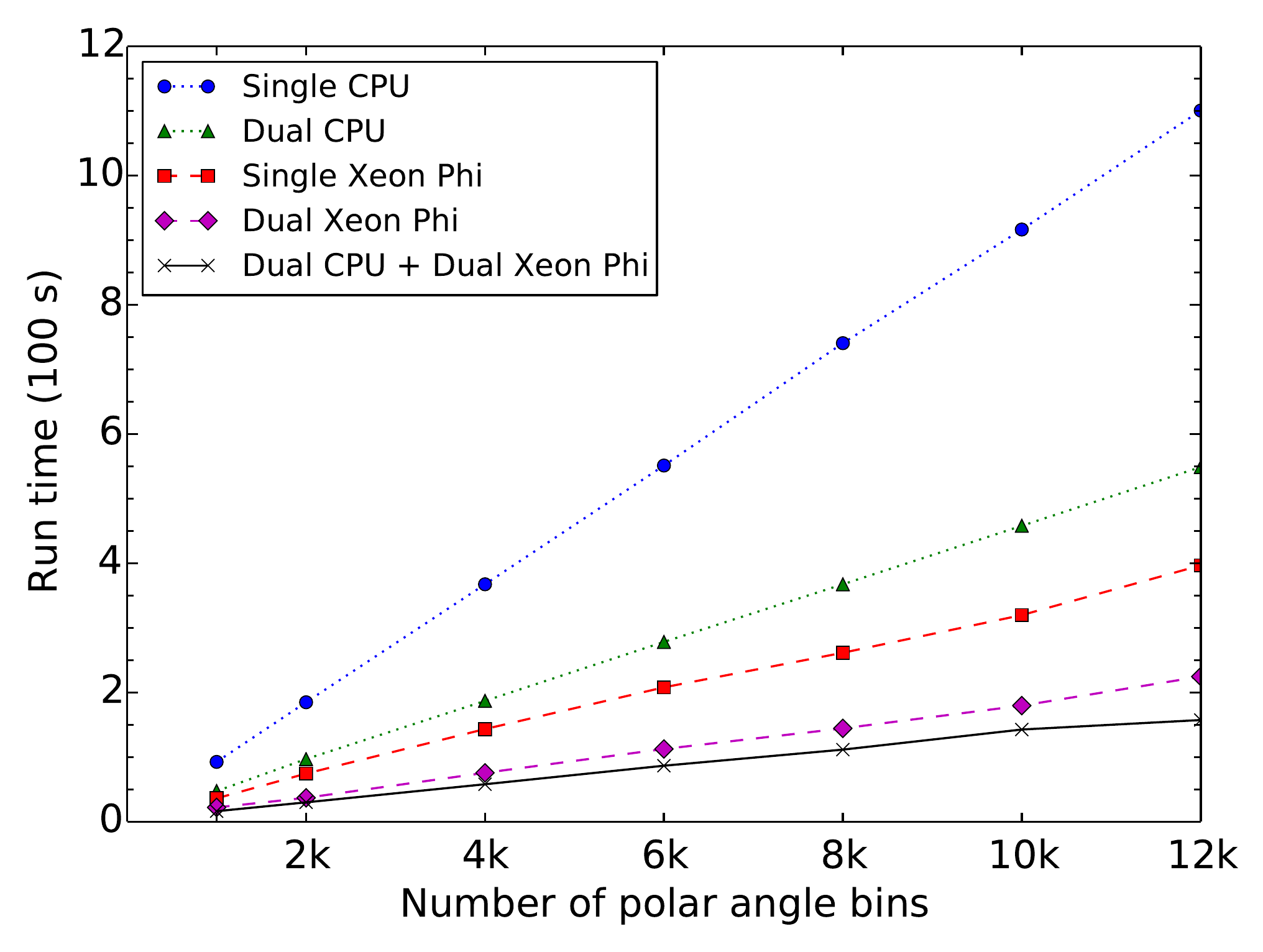}
\caption{Single node XFLAT performance for 1000 radial steps.}
\label{fig:snode}
\end{figure}

We then benchmarked the performance of XFLAT on a single compute node
without I/O. We ran XFLAT on a single CPU, dual CPUs, a single
Xeon Phi, dual Xeon Phis, and both dual CPUs and dual Xeon Phis, respectively.
In all of these tests we used the same input configuration as the ``standard problem''
mentioned previously, except that we varied the
number of polar angle bins from 1,000 up to 12,000 (the memory on a
single Xeon Phi cannot support more than 13,000 polar
angle bins). The results of these tests are shown in Fig.~\ref{fig:snode}.

The run times of these tests scale approximately linearly with problem size, although there are variations in the benchmarks with
Xeon Phis. In single-component tests XFLAT ran 2.5--2.9 times as 
fast on single Xeon Phi as on a single 8-core (Sandy Bridge) Xeon
CPU. In dual-component tests XFLAT ran 2.1--2.6 times as
fast on dual Xeon Phis as on dual CPUs. When both dual CPUs and dual
Xeon Phis were used, XFLAT ran 3.5--3.9 times as fast as on dual CPUs.

\subsection{I/O benchmarks}

\begin{figure}
\centering
\includegraphics[width=\columnwidth]{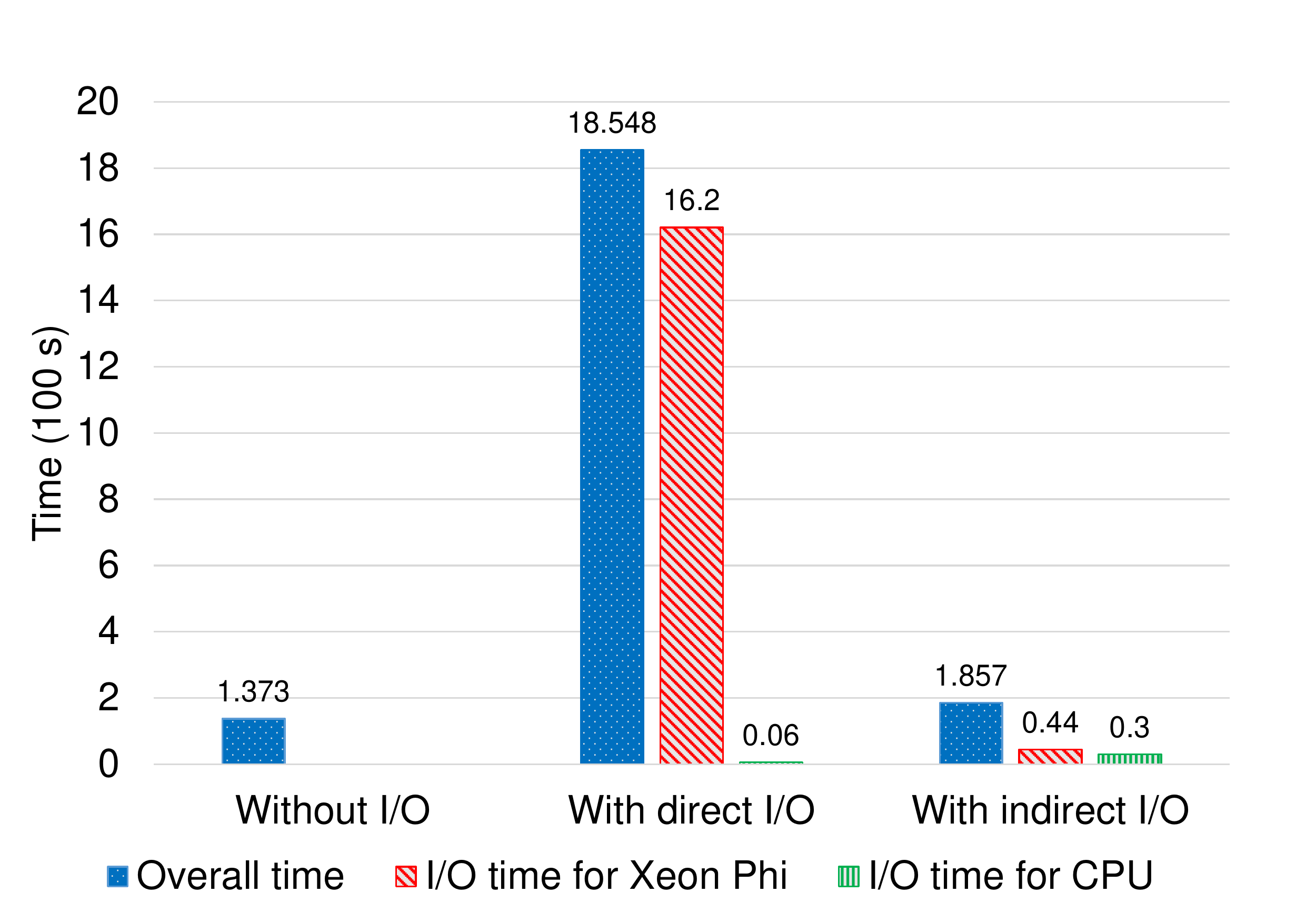}
\caption{Single node I/O performance of XFLAT.}
\label{fig:io}
\end{figure}

The performance of XFLAT can be significantly affected by I/O. 
In its \emph{direct} I/O configuration the XFLAT process on each MPI node,
which can be either a CPU or Xeon Phi,
save its local snapshots of the 
quantum flavor states of neutrinos to the global \texttt{\$SCRATCH}
storage space of Stampede. 
We ran XFLAT with direct I/O and without I/O for
1,000 radial steps on a single node with dual CPUs and dual Xeon
Phis. In the test with direct I/O each MPI node dumps 10
snapshots. The size of a snapshot is 164~MB for a CPU process and 476~MB
for the Xeon Phi because the Xeon Phi process has 2.9 times as many polar
angle bins as the CPU process does. As shown Fig.~\ref{fig:io} XFLAT
ran more than 10 times slower with direct I/O than without I/O.
Most of the extra time in the test with direct I/O was
spent in the I/O module of the Xeon Phi process. 

To mitigate the poor I/O
performance on Xeon Phi, we designed an \emph{indirect} I/O
configuration in which the I/O module of the Xeon Phi process sends
the data to the I/O module of the corresponding CPU
process which then writes to \texttt{\$SCRATCH} for the Xeon Phi. 
We benchmarked XFLAT with indirect I/O, and the result is also shown
in Fig.~\ref{fig:io}. In this test both the overall run time and the
time spent in 
the I/O module of the Xeon Phi process
were significantly reduced compared to that with direct I/O, although
the CPU process had to spend more time in its I/O module to
communicate with the Xeon Phi.
Unfortunately, the indirect I/O configuration breaks the
symmetry between the CPU and Xeon Phi and increases the
complexity of code maintenance.

\subsection{Multi-node benchmarks}

\begin{figure}[t]
\centering
\includegraphics[width=\columnwidth]{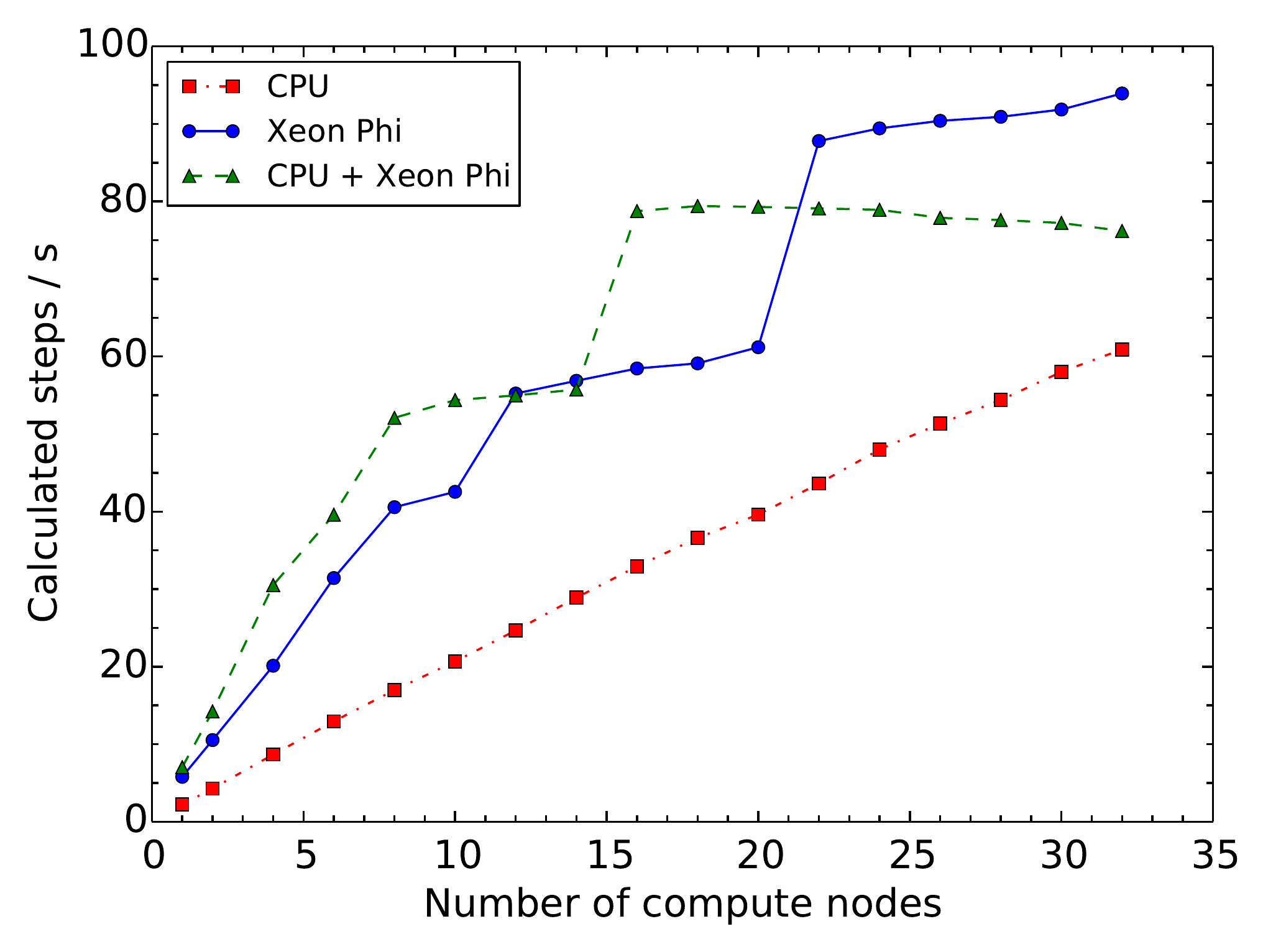}
\caption{Multi-node performance of XFLAT.}
\label{fig:hom}
\end{figure}

We benchmarked XFLAT on multiple compute nodes in the CPU-only, Xeon
Phi-only and hybrid modes which utilize dual CPUs, dual
Xeon Phis and both dual CPUs and dual Xeon Phis for each node, respectively. 
In each of the tests we ran XFLAT for approximately 100 seconds and
computed the number of calculated steps per second. The
results of these benchmarks are shown in Fig.~\ref{fig:hom}.
These results show that the performance of XFLAT scales well in the CPU-only
mode up to 32 nodes, the maximum number of nodes in our
tests. In the Xeon Phi-only mode, however, the performance
scales reasonably well only up to 
12 nodes for the studied problem size, where it
ran 2--2.6 times as fast as in the CPU-only mode. 
In the hybrid mode the performance of
XFLAT scales reasonably well up to 8 nodes where it ran 3--3.5 times as
fast as in the CPU-only mode.

The reason for the poor scaling behavior in the Xeon Phi-only and
hybrid modes is 
because the 244 hardware threads of the Xeon Phi cannot not be fully
utilized for the studied problem size when many compute nodes are used.
For example, in the test with 20 compute nodes in the Xeon Phi-only mode
each Xeon Phi process received
$10,000/40=250$ polar angle bins, which is just a few more than the
244 threads available on the Xeon Phi. As a result, the OpenMP parallel
\texttt{for} loop in 
Fig.~\ref{fig:xflat} iterated twice in each step with most of the
threads idle in the second iteration. 
When 22 compute nodes were used, however, each Xeon Phi process received
227 or 228 polar angle bins, and the computation for these angle bins
was completed in one iteration by employing most of the
threads. The performance of XFLAT decreases when more compute nodes
are used because there is not enough load on each Xeon Phi.

\begin{figure}[t]
\centering
\includegraphics[width=\columnwidth]{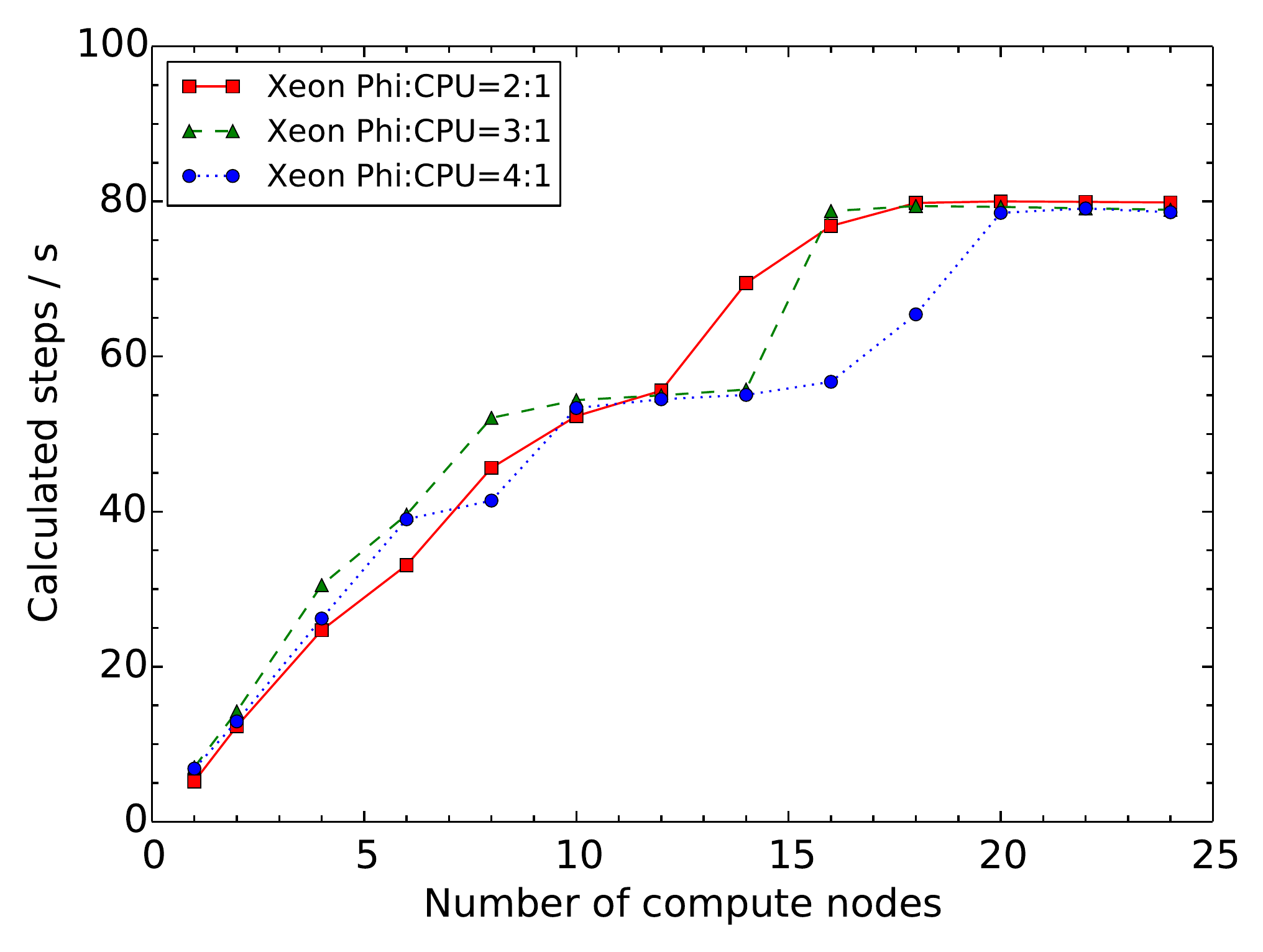}
\caption{Impact of Xeon Phi to CPU load ratio.}
\label{fig:het}
\end{figure}

In all previous hybrid-mode tests we fixed the load ratio between the
Xeon Phi and CPU to be about 3:1. We also benchmarked the multi-node performance
of XFLAT  with different load ratios using the
method described above. In Fig.~\ref{fig:het} we show the results of
three such benchmarks. These results show that, in the regime where
there is enough load on the Xeon Phi, XFLAT generally performs best with three times
as much load on the Xeon Phi as on the CPU. The performance of XFLAT changes
modestly when the Xeon Phi to CPU load ratio is changed from 3:1 to 2:1 or
4:1. Varying the load ratio can have a large impact when the load on
the Xeon Phi is less than is necessary to keep most of its hardware
threads busy for two iterations.  
For example, the test with 14 compute nodes and load ratio 3:1 does not
perform well because each Xeon Phi process has 267 or 268 polar angle
bins, which is again just a few more than 244. Better performance was
achieved by lowering the load ratio and
moving some of the polar angle bins from the Xeon Phi to the
CPU. Indeed, the test
with the 2:1 load ratio (with 238 polar angle bins on each Xeon Phi)
has better performance than that with the 3:1 load ratio.

\section{Conclusions}
We have developed the XFLAT astrophysical simulation
code which fully utilizes all three levels of parallelism available on
Xeon Phi-equipped supercomputers. In general we have found it very helpful
to maintain a single code set which can run efficiently on both the CPU and the
co-processor. 
We benchmarked both the single-node and
multi-node performance of XFLAT in various configurations. 
When there is no I/O involved, each Xeon Phi co-processor can boost
the performance of XFLAT by the equivalent of two or more
8-core (Sandy Bridge) Xeon CPUs.
When there is I/O involved, however, the direct I/O from the Xeon Phi can
reduce the performance of XFLAT substantially. This disadvantage can
be mitigated by redirecting the I/O from the Xeon Phi through the CPU, but a
better fix may need an update in the Xeon Phi and/or its software.
Because the Xeon Phi co-processor has many more hardware threads than the CPU
does, it is essential to have sufficient number of parallel tasks to keep most
of its threads busy. This feature limits the number of compute nodes
for which XFLAT can be run efficiently for the extended neutrino bulb
model. This limitation will be removed with the development and implementation of next-generation multi-dimensional supernova models for neutrino oscillations in XFLAT. 

\section*{Acknowledgments}
This work was supported in part by DOE grant DE-SC0008142 (H.D.) and
NSF grant OCI-1040530 (S.R.A.) at the University of New Mexico. We are
grateful to the Texas Advanced
Computing Center and the UNM Center for
Advanced Research Computing for providing computational resources used
in this work. We thank Dr.\ John Cherry,
Dr.\ Shashank Shalgar, and
Sajad Abbar for their assistance during the development of this
project. 

%
\bibliographystyle{unsrtnat}

\end{document}